\documentclass[12pt]{JHEP3}

\usepackage{amsmath}
\usepackage{amssymb}
\usepackage{amsthm}

\usepackage[all]{xypic}

\newcommand{\be}{\begin{equation}}
\newcommand{\ee}{\end{equation}}

\newcommand{\NN}{\mathbb{N}} 

\newcommand{\hs}[1]{\mbox{hs$[#1]$}}
\newcommand{\hso}[1]{\mbox{hso$[#1]$}}
\newcommand{\W}{\mathcal{W}}
\newcommand{\WD}{\mathcal{WD}}

\DeclareMathOperator{\ve}{\varepsilon}
\DeclareMathOperator{\la}{\Lambda}

\numberwithin{equation}{section}

\def\be{\begin{equation}}
\def\ee{\end{equation}}

\allowdisplaybreaks[3]

\title{\vspace{0.5cm} Minimal Model Holography for SO(2N)}

\author{Matthias R.~Gaberdiel and Carl Vollenweider
\\ ~
\\
Institut f\"ur Theoretische Physik \\
ETH Zurich \\
CH-8093 Z\"urich \\
Switzerland }

\abstract{A duality between the large $N$ 't~Hooft  limit of the $\WD_N$ minimal model CFTs
and a higher spin gravity theory on AdS$_3$ is proposed. The gravity theory has
massless spin fields of all even spins $s=2,4,6,\ldots$, as well as two real scalar fields whose
mass is determined by the 't~Hooft parameter of the CFT. We show that, to leading order in the large
$N$  limit, the 1-loop partition function of the higher spin theory matches precisely with the 
CFT partition function.}

\begin{document}

\section{Introduction}

Recently, a duality between a higher spin gauge theory on AdS$_3$ 
\cite{Prokushkin:1998bq,Prokushkin:1998vn}, and the large $N$ 't~Hooft limit 
of a family of ${\cal W}_N$ minimal models has been proposed \cite{Gaberdiel:2010pz}. This
proposal is the natural 3d/2d analogue of the Klebanov-Polyakov duality
 \cite{Klebanov:2002ja} (see also  \cite{Witten,Mikhailov:2002bp,Sezgin:2002rt} for earlier work), 
relating the  ${\rm O}(N)$ vector model in $d=3$ to a higher spin theory on AdS$_4$
\cite{Vasiliev:2003ev} (see for example
\cite{Vasiliev:1999ba,Bekaert:2005vh,Iazeolla:2008bp,Campoleoni:2009je} for reviews);
for recent progress with the Klebanov-Polyakov proposal see 
\cite{Giombi:2009wh,Giombi:2010vg,Koch:2010cy,Giombi:2011ya}.

In the proposal of \cite{Gaberdiel:2010pz} the higher spin gravity theory on AdS$_3$ 
has an infinite tower of massless fields with spins $s=2,3,\dots$, coupled to two complex scalars. 
The interactions among the higher spin fields are dictated by the higher spin Lie algebra
\hs{\lambda},  whose commutation relations depend on a free parameter $\lambda$.  
The scalars have equal mass  fixed by 
the algebra \cite{Vasiliev:1992ix,Prokushkin:1998bq,Prokushkin:1998vn},
\be\label{mass}
M^2 = -1 + \lambda^2 \ ,
\ee
but they are quantised with opposite (conformally invariant) boundary conditions, and 
the corresponding boundary fields have conformal weights
\be\label{hpm}
h_+ = \frac{1+\lambda}{2} \ , \qquad h_- = \frac{1-\lambda}{2} \ .
\ee
The dual CFT is the large $N$ 't~Hooft limit of the ${\cal W}_N$ minimal models. The 
$\W_N$ minimal models can be described as cosets (see \cite{Bouwknegt:1992wg} for a review)
\be
\frac{\mathfrak{su}(N)_k \oplus \mathfrak{su}(N)_1 }{ \mathfrak{su}(N)_{k+1}} \ .
\end{equation}
The large $N$ limit is taken while holding the 't~Hooft coupling
\be
\lambda = \frac{N}{N+k} 
\ee
fixed, where $\lambda$ is to be identified with the parameter appearing in the commutation relations
of the higher spin algebra.

The proposal of \cite{Gaberdiel:2010pz} has been tested in a variety of ways: in particular,
it was shown that the leading large $N$ partition function of the family of CFTs  agrees with
the 1-loop determinant  of the bulk theory \cite{GGHR}, and that the asymptotic symmetries of the 
higher spin gravity theory match those of the CFT \cite{Gaberdiel:2011wb}, extending the earlier analysis of 
\cite{Henneaux:2010xg,Campoleoni:2010zq,Gaberdiel:2010ar}. Finally, the known RG flows
between $\W_N$ models of different $k$ were found to agree with the expected behaviour in
the bulk~\cite{Gaberdiel:2010pz}.
\medskip

In this paper we propose another variant of this duality. The higher spin gravity theory on AdS$_3$
can be consistently truncated to the massless gauge fields with even spin $s=2,4,6,\ldots$.
We will couple this higher spin theory to two {\em real scalar fields}, whose
mass is again fixed by the algebra precisely as in (\ref{mass}) \cite{Prokushkin:1998bq}. 
The two real scalar fields are quantised with opposite (conformally invariant) boundary conditions and 
therefore the corresponding conformal weights are also given by (\ref{hpm}).

The dual conformal field theory is the 't~Hooft limit of the $\WD_N$ series. The $\WD_N$ minimal
models can be described by the cosets 
\be\label{socos}
\frac{\mathfrak{so}(2N)_k \oplus \mathfrak{so}(2N)_1 }{ \mathfrak{so}(2N)_{k+1}} \ ,
\end{equation}
and the large $N$ limit is now taken with constant 't~Hooft coupling
\be\label{tHooft}
\lambda = \frac{2N}{2N+k-2} \ .
\ee
In this paper we give substantial evidence in favour of this proposal. In particular, we show that the
leading large $N$ partition function of the $\WD_N$ minimal models  agrees precisely with
the 1-loop determinant  of the bulk theory. 
Many of our arguments are fairly similar to \cite{GGHR}, but there are also some new phenomena.
For example, the spinor representations of $\mathfrak{so}(2N)$ decouple in the large $N$ 't~Hooft 
limit since their conformal dimensions go to infinity. The representations that are of interest
are therefore the vector (and tensor) representations of $\mathfrak{so}(2N)$, and their branching functions 
behave  essentially like those of the tensor powers of the fundamental (or anti-fundamental) 
representation of $\mathfrak{su}(N)$. 
However, unlike the fundamental representation of $\mathfrak{su}(N)$, the vector representation
of $\mathfrak{so}(2N)$ is self-conjugate, and thus there is no separate conjugate representation.
This mirrors the fact that the two scalar fields in AdS$_3$ are real (rather than complex)
scalar fields.
\medskip

The paper is organised as follows. In Section~\ref{grav} we explain the calculation of the 
bulk 1-loop determinant, and show that the answer can be written in terms of  ${\rm U}(\infty)$ characters; 
this analysis follows very straightforwardly from \cite{GGHR}. The core of the paper is Section~\ref{CFT},
where we explain how to calculate the 't~Hooft limit of the CFT partition function for the
$\WD_N$ minimal models. As in \cite{GGHR}, the large $N$ limit is subtle in that certain states 
decouple from the correlation functions, and thus do not contribute to the partition function as
$N\rightarrow \infty$; once this is taken into account, the resulting partition function matches exactly
the bulk answer. Finally, we comment on the structure of the bulk higher spin theory in 
Section~\ref{concl}. 
Some of the details of the Lie algebraic analysis is described in Appendix~\ref{so},
and the fusion calculation justifying the decoupling statement is outlined in Appendix~\ref{fusion}.
\bigskip

\noindent
{\bf Note:} As this paper was written up,  \cite{Ahn:2011pv} appeared that has some overlap
with the present paper. He also proposes the duality we describe (giving as evidence the matching
of the RG flows), but does not confirm the identity of partition functions.

\section{The gravity partition function}\label{grav}

Let us begin by calculating the 1-loop contribution of the bulk degrees of freedom. Using
the techniques of \cite{David:2009xg} the contribution of a massless spin $s$ field
to the 1-loop partition function on thermal AdS$_3$ was found to be \cite{Gaberdiel:2010ar}
\be
Z_{{\rm spin}\ s} = \prod_{n=s}^{\infty} \frac{1}{|1-q^n|^2} \ . 
\ee
Combining the contributions from the fields of even spin $s=2,4,6,\ldots$, the higher spin
part of the 1-loop partition function therefore equals
\be\label{MMe}
Z_{\rm hs} = \prod_{\substack{s = 2 \\ s \, \mathrm{even}}}^\infty \; \prod_{n = s}^\infty 
\frac{1}{{| 1 - q^n |}^2} \equiv | M^e (q) |^2\ ,
\end{equation}
where $M^e$ is the modified `even'  MacMahon function. 

The contribution of a real scalar field to the 1-loop partition function was determined in 
\cite{Giombi:2008vd} (see also \cite{David:2009xg}) to be
\begin{equation}
Z_{\mathrm{scal}} (h) = \prod^\infty_{j,j'=0} \frac{1}{1 - q^{h+j} \bar{q}^{h + j'}} \ ,
\end{equation}
where $h$ is the conformal dimension of the corresponding boundary field. 
For $h=h_\pm$ as in (\ref{hpm}) it was shown in \cite{GGHR} that the scalar 1-loop partition 
function can be written as 
\be\label{scalY}
Z_{\mathrm{scal}} (h_\pm) = \sum_Y |P_Y^\pm(q) |^2 \ , 
\ee
where $P_Y(q)$ are the (specialised) Schur functions
\begin{eqnarray}\label{Schur}
P_Y(q) & = & \chi_Y^{\mathfrak{u}(\infty)}(z_i) \ , \qquad z_i = q^{i-\frac{1}{2}} \nonumber \\
P_Y^\pm(q) & = & q^{\pm\frac{\lambda}{2} B(Y)} \, P_Y(q) \ .
\end{eqnarray}
Here $Y$ labels Young tableaux, and the sum  in (\ref{scalY}) runs over all Young tableaux $Y$
whose number of boxes $B(Y)$ is finite, see \cite{GGHR} for further details.

Combining these 1-loop contributions with the classical action $Z_{\rm cl} = (q\bar{q})^{-\frac{c}{24}}$
the total partition function of the bulk theory equals
\be\label{bulka}
Z_{\rm bulk} = (q\bar{q})^{-\frac{c}{24}} \, |M^e(q)|^2 \, \sum_{Y_+,Y_-} 
| P_{Y_+}^+(q)|^2 \cdot | P_{Y_-}^-(q)|^2 \ .
\ee
In the following we want to reproduce  this partition function from  the
't~Hooft limit of the coset CFTs (\ref{socos}).

\section{The CFT analysis}\label{CFT}

The conformal field theory we are interested in is the coset model
\begin{equation}
\frac{\mathfrak{so}(2N)_k \oplus \mathfrak{so}(2N)_1} {\mathfrak{so}(2N)_{k + 1}} \ ,
\end{equation}
whose central charge equals 
\begin{equation}\label{central}
\begin{split}
c\equiv c_N(p) &= \mathrm{dim} \; \mathfrak{so}(2N) \left[ \frac{k}{k + h^{\vee}} + \frac{1}{1 + h^{\vee}} 
- \frac{k+1}{k + 1 + h^{\vee}}  \right] \\
&= N \left[ 1 - \frac{(2N -1) (2N - 2)}{p (p+1)} \right] \ ,
\end{split}
\end{equation}
where  $h^{\vee}=2N-2$ is the dual Coxeter number of the finite dimensional
Lie algebra $\mathfrak{so}(2N)$
of dimension $\mathrm{dim} \; \mathfrak{so}(2N) = N (2N -1)$,  and 
$p \equiv k+2N  -2$. The highest weight representations (hwr) of the coset are labelled by
triplets $(\Lambda_+,\mu;\Lambda_-)$, where $\la_+$ and $\la_-$ are integrable hwr of $\mathfrak{so}(2N)_k$
and $\mathfrak{so}(2N)_{k + 1}$, respectively, while $\mu$ is a $\mathfrak{so}(2N)_{1}$ hwr. The triplets have
to satisfy the selection rule that $\la_+ + \mu - \la_-$ (interpreted as a weight of the finite
dimensional Lie algebra $\mathfrak{so}(2N)$) lies in the root lattice of $\mathfrak{so}(2N)$.
Modulo the root lattice, the weight lattice of $\mathfrak{so}(2N)$ has four conjugacy classes, and  there is 
precisely one level one representation in each conjugacy class; thus the selection rule determines $\mu$ 
uniquely, and we can therefore label our coset representations by the pairs 
$(\Lambda_+;\Lambda_-)$. In addition there is the field identification $(\la_+;\la_-)\cong (A\la_+;A\la_-)$,
where $A$ is the outer automorphism of the affine algebra $\mathfrak{so}(2N)_k$ and 
$\mathfrak{so}(2N)_{k+1}$, respectively. $A$ permutes the four roots of the extended Dynkin diagram with 
Kac label $1$. In the 't~Hooft limit, this field identification does not lead to any identifications among
the representations that are of interest to us, and we can thus ignore it.

The conformal weights of the hwr $(\Lambda_+;\Lambda_-)$ can be calculated from the 
coset description as 
\begin{equation}
h_{(\la_+; \la_-)} =  \frac{C_N(\la_+)}{p} + \frac{C_N(\mu)}{2 N - 1} - \frac{C_N(\la_-)}{p + 1} + n \ , 
\end{equation}
where $C_N(\la)$ is the quadratic Casimir that is defined in Appendix~\ref{so}, while 
$n$ denotes the conformal weight above the ground state at which $\la_-$ appears in the representation 
$\la_+ \oplus \mu$. Alternatively, we may use the Drinfeld-Sokolov description of the $\WD_N$
model (see {\it e.g.}\ \cite[section~6]{Bouwknegt:1992wg}), in terms of which the conformal weight
of the hwr $(\Lambda_+;\Lambda_-)$ equals
\begin{equation}\label{hco}
h_{( \Lambda_{+}; \Lambda_- )} = 
\frac{c_N(p)}{24} - \frac{N}{24} 
+ \frac{1}{2 p (p+1)} \bigl| (p +1) (\Lambda_+ + \rho ) - p ( \Lambda_- + \rho) \bigr|^2 \ ,
\end{equation}
where $\rho$ denotes the Weyl vector of the finite dimensional Lie algebra $\mathfrak{so}(2N)$.
Using the Lie algebraic identities
\begin{equation}
\mathrm{dim} \; \mathfrak{g} = ( \hbox{rank}\; \mathfrak{g})\,  (1 + h^{\vee}) \ , \qquad
\frac{\rho^2}{2 h^{\vee}} = \frac{\mathrm{dim} \: \mathfrak{g}}{24} \ , 
\end{equation}
for $\mathfrak{g} = \mathfrak{so}(2N)$ with ${\rm rank}(\mathfrak{so}(2N))=N$, (\ref{hco}) can be rewritten 
as 
\begin{equation}
h_{( \Lambda_{+}; \Lambda_- )} = 
\frac{1}{2 p (p+1)} \Bigl( \bigl| (p +1) (\Lambda_+ + \rho ) - p ( \Lambda_- + \rho) \bigr|^2 - \rho^2 \Bigr) \ .
\end{equation}
For example, for the representations where either $\la_+$ or $\la_-$ is the trivial representation, we have 
\begin{eqnarray}
h_{( 0; \Lambda_- )} & = &  \frac{1}{2 (p+1)} \left(  p ( \Lambda_-)^2 
- 2  \langle \Lambda_-, \rho \rangle \right) = \frac{1}{2} (\Lambda_-)^2 - \frac{1}{(p+1)} C_N(\Lambda_-)
\nonumber \\[4pt] 
h_{( \Lambda_+; 0 )} & = & \frac{1}{2 p} \left(  (p + 1) ( \Lambda_+)^2 
+ 2 \langle \Lambda_+, \rho \rangle \right)  
= \frac{1}{2} (\Lambda_+)^2 + \frac{1}{p} C_N(\Lambda_+)\ .
\end{eqnarray}

\subsection{The 't~Hooft limit}

We are interested in the 't~Hooft limit where we take $N$ and $k$ to infinity, keeping the ratio 
\begin{equation}
\lambda = \frac{2N}{k+2N-2} = \frac{2N}{p} 
\end{equation}
fixed. It follows from the analysis of Appendix~\ref{so}, see (\ref{A.15}), that for representations 
$\la$ whose Dynkin labels satisfy $\Lambda_{N-j}\neq 0$, where we keep $j$ fixed as we take the
large $N$ limit, we have $\la^2\sim N$ and $C_N(\la)\sim~N^2$. Thus the conformal weight of 
the corresponding coset representations $(\la;0)$ or $(0;\la)$ will be proportional to $N$, and these
states decouple in the 't~Hooft limit. For example, for the two spinor representations 
$s=[0^{N-2},1,0]$ and $c=[0^{N-1},1]$, we have from (\ref{A.13})
\begin{equation}
h_{(s;0)} = h_{(c;0)} = \tfrac{N}{8} \bigl( 1 + \tfrac{2N-1}{p} \bigr) \ , \qquad
h_{(0;s)} = h_{(0;c)} = \tfrac{N}{8} \bigl( 1 - \tfrac{2N-1}{p+1} \bigr) \ .
\end{equation}
Thus we are only interested in representations for which $\Lambda_i=0$ for
$i\geq j$, where $j$ is kept fixed as we take $N\rightarrow \infty$. These representations
are precisely those that appear in finite tensor powers of the vector representation
$v=[1,0^{N-1}]$. Note that for the vector representation we find on the other hand
\begin{equation}\label{hvec}
h_{(v;0)} =   \tfrac{1}{2} \bigl( 1 + \tfrac{2N-1}{p} \bigr) \cong \tfrac{1}{2} (1 + \lambda) \ , \qquad
h_{(0;v)}  =   \tfrac{1}{2} \bigl( 1 - \tfrac{2N-1}{p+1} \bigr) \cong \tfrac{1}{2} (1 - \lambda) \ , 
\end{equation}
where we have denoted by $\cong$ the value in the 't~Hooft limit; this matches precisely
(\ref{hpm}) from above.

As is explained in Appendix~\ref{so}, the representations with $\Lambda_{N-1}=\Lambda_N=0$
can be labelled by Young tableaux, and the condition that  $\Lambda_i=0$ for $i\geq j$ with $j$
fixed means that the corresponding Young tableaux have only finitely many boxes.
The situation is therefore similar to what happened in \cite{Gaberdiel:2010pz}, and the 
state $(0;v)$ plays a similar role to that of $(0;{\rm f})$ there, and likewise for $(v;0)$. 
However, there is one important difference: the vector representation of
$\mathfrak{so}(2N)$ is its own conjugate representation, and thus there is no analogue
of $(0;\bar{\rm f})$ or  $(\bar{\rm f};0)$ in the current context. This mirrors the 
fact that the corresponding scalar in the bulk is a real scalar (rather than a complex 
scalar as in \cite{Gaberdiel:2010pz}).

One consequence of this  reality property is that the tensor product rules of the 
$\mathfrak{so}(2N)$ representations differ from the usual Young tableaux rules. For example, 
the tensor product of  $v$ with itself equals
\begin{equation}\label{tensorv}
[1,0^{N-1}] \otimes [1,0^{N-1}] = [2, 0^{N-1}] \oplus [0,1,0^{N-2}] \oplus [0^{N}] \ ,
\end{equation}
and thus contains the trivial representation, $[0^{N}]$, as well as the adjoint
representation adj$=[0,1,0^{N-2}]$.

\subsection{The branching functions}

Next we want to derive a formula for the characters of the corresponding coset representations.
The argument follows again closely \cite{GGHR}, but the details of the calculation are different. 

At finite $N$ and $k$, the character of the hwr $(\la_+; \la_-)$ is given by (see {\it e.g.}\ 
\cite[eq.\ (7.51)]{Bouwknegt:1992wg})
\begin{equation} \label{braFu}
b_{( \Lambda_{+}; \Lambda_- )} (q) = \frac{1}{\eta(q)^N} 
\sum_{w \in \hat{W}} \epsilon(w) q^{\frac{1}{2 p (p +1)} \bigl( (p+1) w( \la_+ + \rho) - p (\la_{-} + \rho) \bigr)^2},
\end{equation}
where $\hat{W}$ denotes the affine Weyl group. The affine Weyl group is 
isomorphic to the semidirect product of the finite Weyl group times translations by elements of the 
(co-)root lattice of the finite dimensional Lie algebra $\mathfrak{so}(2N)$. Its action on the weight 
$\la + \rho$ is given by
\begin{equation} \label{aWeylGroup}
w (\la + \rho) = w_{\mathrm{finite}} (\la + \rho) + (k+2N-2) \mathbf{P}\ ,
\end{equation}
where $w_{\mathrm{finite}}$ is a finite Weyl reflection and $\mathbf{P}$ is an appropriate co-root. 
Note that the term involving $\mathbf{P}$ in eq. (\ref{aWeylGroup}) is proportional to $p=k + 2N-2$. In the 
large $N$ 't~Hooft limit the corresponding terms therefore do not contribute. Thus we may restrict
the sum in eq.~(\ref{braFu}) to the finite Weyl group $W$
\begin{equation} \label{finiteW}
b_{( \Lambda_{+}; \Lambda_- )} (q) 
\cong \frac{1}{\eta(q)^N} \sum_{w \in W} \epsilon(w) q^{\frac{1}{2 p (p +1)} 
\bigl( (p+1) w( \la_+ + \rho) - p (\la_- + \rho) \bigr)^2}\ .
\end{equation}
Here the \mbox{symbol $\, \cong \,$} indicates that this equality is only true in the `t~Hooft limit. 
Since the elements of the finite Weyl group leave the norm of weight vectors invariant, 
the expansion of the exponent of $q$ in eq.~(\ref{finiteW}) yields
\begin{equation} \label{qexp}
\left( 1 + \tfrac{1}{2 p (p+1)} \right) \rho^2 + \left( 1 + \tfrac{1}{p} \right) C_N(\la_+) 
+ \bigl( 1 - \tfrac{1}{p+1} \bigr) C_N(\la_-) 
- \langle w (\la_+ + \rho), \la_- + \rho \rangle \ ,
\end{equation}
where $C_N (\la) = \frac{1}{2} \left( \la^2 + 2 \langle \la, \rho \rangle \right)$ is the 
quadratic Casimir of the re\-pre\-sen\-ta\-tion $\la$. For the representations with 
$\la_{N-1}=\la_{N}=0$,
the quadratic Casimir has the expansion (see eq.~(\ref{A.18}))
\begin{equation}
C_N(\la) = B (\la) \left(N - \tfrac{1}{2} \right) + \tfrac{1}{2} D(\la) \ ,
\end{equation} 
where $B (\la)$ is the total number of boxes in the Young tableau corresponding to $\la$, 
and $D(\la)$ is defined in (\ref{A.17}). In the 't~Hooft limit, eq.~(\ref{qexp}) therefore becomes
\begin{equation} \label{qexp2}
\left( 1 + \tfrac{1}{2 p (p+1)} \right) \rho^2 + C_N(\la_+) + \tfrac{\lambda}{2} B(\la_+) 
+ C_N(\la_-) - \tfrac{\lambda}{2} B(\la_-) - \langle w (\la_+ + \rho), \la_- + \rho \rangle \ .
\end{equation}
Inserting this  into eq.~(\ref{finiteW}) we obtain
\begin{equation} \label{prae}
b_{( \Lambda_{+}; \Lambda_- )} (q) \cong 
q^{\rho^2 + \tfrac{1}{2 p (p+1)} \rho^2} \frac{q^{C_N(\la_+) + C_N(\la_-)} 
q^{\frac{\lambda}{2}(B(\la_+) - B(\la_-))}}{\eta(q)^N} \sum_{w \in W} \epsilon(w) \,
q^{- \langle w(\la_+ + \rho), \la_- + \rho \rangle}\ .
\end{equation}
Up to now, we acted as though the branching functions would converge in the 't~Hooft limit. However, 
this is not true because the characters are all proportional to $q^{-c/24}$, and since in the 't~Hooft limit
the central charge, see eq.~(\ref{central}),  diverges as 
\begin{equation}
c \cong N (1-\lambda^2) \ , 
\end{equation}
this will lead to a divergence. In order to extract
this divergent part, we note that 
\begin{equation}
\frac{c}{24} = \frac{N}{24} - \frac{\rho^2}{2 p (p+1)} \ , 
\end{equation}
where we have rewritten (\ref{central}) using (\ref{Weylsq}). Thus we can rewrite eq.~(\ref{prae}) as 
\begin{equation} \label{bff}
q^{\frac{c}{24}} b_{( \Lambda_{+}; \Lambda_- )} (q) \cong  
\frac{q^{C_N(\la_+) + C_N(\la_-)} 
q^{\frac{\lambda}{2}(B(\la_+) - B(\la_-))}}{\tilde{\eta}(q)^N} \, 
q^{\rho^2} \sum_{w \in W} \epsilon(w) \, q^{- \langle w(\la_+ + \rho), (\la_- + \rho) \rangle}\ ,
\end{equation}
where $\tilde{\eta}(q)$ is the modified eta function without the factor of $q^{\frac{1}{24}}$, {\it i.e.}
\be
\tilde{\eta}(q) = \prod_{n=1}^{\infty} (1-q^n) \ . 
\ee
Next we use the Weyl denominator formula eq.\ (\ref{Weylso}) for $\mathfrak{so}(2N)$, to obtain 
the identity 
\begin{eqnarray}
\frac{q^{\rho^2}}{\tilde{\eta}(q)^N} \,  \sum_{w \in W} \epsilon(w) q^{ - \langle w(\rho), \rho \rangle} 
& = & \frac{1}{\tilde{\eta}(q)^N} \, \prod_{n=1}^{2N-3} (1 - q^n)^{N- \left \lceil \frac{n - 1}{2} \right \rceil} 
\prod_{n=N}^{2N-3} (1 - q^n)^{-1} \nonumber \\
& \cong &  \prod_{\substack{s = 2 \\ s \, \mathrm{even}}}^\infty 
\prod_{n = s}^\infty \frac{1}{1 - q^n} = M^{\rm e}(q) \ , 
\end{eqnarray}
where $M^{\rm e}(q)$ is the modified even MacMahon function introduced before
in (\ref{MMe}). 

\noindent Inserting the above identity into  eq.~(\ref{bff}), we thus conclude that 
\begin{equation} \label{bfff}
b_{( \Lambda_{+}; \Lambda_- )} (q) \cong q^{-\frac{c}{24}}  \, M^e (q)  \, 
q^{C_N(\la_+) + C_N(\la_-)} \, q^{\frac{\lambda}{2}(B(\la_+) - B(\la_-))} \, \frac{S_{\Lambda_+ \Lambda_-}}{S_{00}} \ ,
\end{equation}
where we have used that (see {\it e.g.}\ \cite[eq.\ (2.7.24)]{Fuchs})
\be\label{Siden}
\frac{\sum_{w \in W} \epsilon(w) q^{- \langle w(\la_+ + \rho), (\la_- + \rho) \rangle}}
{\sum_{w \in W} \epsilon(w) q^{ - \langle w(\rho), \rho \rangle}} = \frac{S_{\Lambda_+ \Lambda_-}}{S_{00}} \ ,
\ee
with $S_{\la_+\la_-}$ the modular $S$-matrix of $\mathfrak{so}(2N)$ at level $\hat{k}$ with
$q=\exp(\tfrac{2\pi i }{ \hat{k} + 2N -2 } )$. Following the same arguments as in \cite{GGHR}
we have
\be
q^{C_N(\la_+) + C_N(\la_-)}\, \frac{S_{\Lambda_+ \Lambda_-}}{S_{00}}  = \sum_{\la} N_{\la_+ \la_-}{}^{\la}\, 
q^{C_N(\la)}\, \frac{S_{\Lambda 0}}{S_{00}} \ ,
\ee
where $N_{\la_+ \la_-}{}^{\la}$ denote the fusion rules (or Clebsch-Gordon) coefficients of 
$\mathfrak{so}(2N)$.  Using (\ref{Siden}), the last factor can be written as 
\begin{eqnarray}
q^{C_N(\la)}\, \frac{S_{\Lambda 0}}{S_{00}}  & =  &
q^{C_N(\la)}\, \frac{\sum_{w \in W} \epsilon(w) q^{- \langle w(\la + \rho), \rho \rangle}}
{\sum_{w \in W} \epsilon(w) q^{ - \langle w(\rho), \rho \rangle}}  \nonumber \\
& = & q^{\frac{1}{2}\sum_i l_i^2}\, 
\prod_{i = 2}^N \prod_{j = 1}^{i - 1} \frac{ \left( 1 - q^{l_j - l_i + i - j} \right)}{\left( 1 - q^{i - j} \right)}
\prod_{i = 2}^N \prod_{j = 1}^{i - 1} \frac{1 - q^{l_j + l_i + 2N - i - j}}{1 - q^{2N - i - j}} \ .
\end{eqnarray}
In the second line we have used eqs.~(\ref{nom}) and (\ref{denom}) from Appendix~\ref{so}, as
well as the explicit formula for the quadratic Casimir $C_N(\la)$, see eq.~(\ref{quadC}). 
Here $l_j$ denotes the number of boxes in the $j^{\rm th}$ row of the Young tableau
$Y(\Lambda)$ associated to $\Lambda$.
In our limit, $l_j$ and $l_i$ are independent of $N$, and thus the second product converges to 
$1$ as $N \rightarrow \infty$. Hence the expression simplifies to 
\begin{equation}
q^{C_N(\la)}\, \frac{S_{\Lambda 0}}{S_{00}}  \cong 
q^{\frac{1}{2} \sum_i l_i^2}
\prod_{i = 2}^N \prod_{j = 1}^{i - 1} \frac{ \left( 1 - q^{l_j - l_i + i - j} \right)}{\left( 1 - q^{i - j} \right)} =
q^{\frac{1}{2} \sum_{i} l_i^2 }  \prod_{(i j) \in Y(\Lambda)} \left( 1 - q^{h_{ij}} \right) \ ,
\end{equation}
where the last product runs over all boxes of the Young tableau $Y(\Lambda)$, 
and $h_{ij}$ is the hook length of the box labelled by $(ij)$. (The last identity is the usual hook 
length formula, see {\it e.g.}\ \cite[Chapter 3]{MacDonald}.) Finally, we recall from \cite{GGHR} that
with $z_i=q^{i-\frac{1}{2}}$, 
\begin{equation} \label{su}
\chi_\Lambda^{\mathfrak{u}(N)}(z_i) \cong q^{\frac{1}{2} \sum_{j = 1}^N c_j^2 } 
\prod_{(i j) \in Y(\Lambda)} \left( 1 - q^{h_{ij}} \right) = P_{Y(\Lambda)}(q)\ ,
\end{equation}
where $P_Y(q)$ is defined in (\ref{Schur}), and $c_i$ denotes the number of boxes in the $i^{\rm th}$
column. 
Combining the last two equations we therefore conclude that 
\be\label{generalBra}
b_{( \Lambda_{+}; \Lambda_- )} (q) \cong q^{-\frac{c}{24}}\, M^e (q)  \, 
q^{\frac{\lambda}{2}(B(\la_+) - B(\la_-))} 
\sum_{\la} N_{\la_+ \la_-}{}^{\la} \, P_{Y^T(\la)}(q) \ ,
 \ee
where $Y^T$ is the transposed Young tableau of $Y$, where the roles of the rows and columns
(and thus of $l_i$ and $c_i$) have been interchanged.

\subsection{Matching the partition functions}

As in \cite{GGHR}, (\ref{generalBra}) is the limit of the coset branching functions.
However, as explained in some detail there, it does not describe
the correct CFT partition function in the 't~Hooft limit because certain states become null
and decouple. As an example consider the case $\Lambda_+=\Lambda_-=v$ for 
which it follows from (\ref{tensorv}) that 
\be
b_{(v;v)}(q) \cong q^{-\frac{c}{24}} M^e(q) \Bigl( P_0(q) + P_{\rm adj}(q) + P_{[2,0^{N-1}]}(q) \Bigr) \ .
\ee
On the other hand, we can analyse the representation 
$(v;v)\equiv (v;0)\otimes (0;v)$, repeating the fusion calculation of \cite{GGHR}. The 
analysis is essentially identical to \cite{GGHR}, except that now we cannot use the $W$-modes 
of spin $s=3$ in order to obtain constraints on the fusion product, but only the $L$ and $U$-modes 
of spin $2$ and $4$, respectively. We have checked (see Appendix~\ref{fusion} for details) 
that the analysis  goes through essentially unchanged, and that the resulting representation
is exactly as in  \cite{GGHR} (see eq.\ (2.20) of that paper) 
\be\label{vvdia}
 (v;v): \qquad 
\xymatrix@C=1pc@R=2.2pc{
   & \vdots & & \vdots & & \vdots & & \vdots \\
   & 2 & & \ar@<0.4ex>[dr]^{L_1} \rho & & \ar@<-0.4ex>[dl] \xi & & \ar@<-0.4ex>[lldd]_{L_2} T \\
  & 1 & & & \psi\ar@<0.4ex>[ul]^{L_{-1}} \ar@<-0.4ex> [ur] \ar[dr]_{L_1}& & &\\
 L_0 =& 0 & &  & & \omega\ar@<-0.4ex>[uurr]_{L_{-2}} & & 
}
\ee 
Thus the states corresponding to $P_0(q)$ above --- these are the states generated 
from $\omega$ --- become null, and have to be removed
from the spectrum. Taking this into account the actual CFT character of the limit theory equals then
\begin{eqnarray}
{\rm ch}^{\rm CFT}_{(v;v)} (q) & \cong & q^{-\frac{c}{24}} M^e(q) \,
\Bigl( P_{\rm adj}(q) + P_{[2,0^{N-1}]}(q) \Bigr)
\nonumber \\
& \cong & q^{-\frac{c}{24}} M^e(q)\,   \bigl( P_{v}(q)  \bigr)^2 \ .
\end{eqnarray}
Extrapolating this to the general case, we conclude, as in \cite{GGHR}, 
that we should restrict the sum in eq.~(\ref{generalBra}) to those $\Lambda$ for which
$B(\la) = B(\la_+) + B(\la_-)$. With this modification, 
eq.~(\ref{generalBra}) becomes then
\begin{equation}\label{3.34}
{\rm ch}^{\rm CFT}_{(\Lambda_+; \la_-)} (q) \cong q^{-\frac{c}{24}} \, M^e (q) \,
P^+_{Y_+^T} (q) \cdot  P^-_{Y_-^T} (q) \ ,
\end{equation}
where $Y_\pm=Y(\Lambda_\pm)$, and 
the $\lambda$-dependent prefactors have been absorbed into the definition of 
$P^{\pm}_{Y_{\pm}} (q)$, see (\ref{Schur}). The total CFT partition function is then
\be\label{bounda}
Z^{\rm CFT}(q,\bar{q}) = (q\bar{q})^{-\frac{c}{24}}  \, |M^e(q)|^2 \, 
\sum_{Y_+ Y_-} |P^+_{Y_+^T} (q)|^2  \cdot  | P^-_{Y_-^T} (q)  |^2 
\ee
in perfect agreement with the bulk partition function (\ref{bulka}).

\section{Comments}\label{concl}

In this paper we have shown that the CFT partition function of the $\WD_N$ models in the 
't~Hooft limit (\ref{tHooft}) agrees precisely with the 1-loop determinant of a higher spin gravity theory
in AdS$_3$. The higher spin theory contains massless fields of even spin $s=2,4,6,\ldots$,
as well as two real massive scalar fields whose mass is related to the 't~Hooft parameter
of the CFT limit as in eq.~(\ref{mass}). Our analysis is very similar to \cite{GGHR}, and gives
convincing evidence for the duality.

One may ask whether we can specify the bulk gravity theory further. For the duality proposed
in \cite{Gaberdiel:2010pz}, it was argued convincingly \cite{Gaberdiel:2011wb} (see
also Section~6 of \cite{GGHR}) that the higher spin part of the bulk theory is described by a
Chern-Simons theory based on the Lie algebra \hs{\lambda}. This Lie algebra can be defined
via (see {\it e.g.}\ \cite{Gaberdiel:2011wb} for details)
\be\label{Usl2}
{\bf 1} \oplus \hs{\lambda} \cong \frac{U(\mathfrak{sl}(2))}{\langle C_2 - \mu {\bf 1} \rangle} \ , \qquad
\mu = \tfrac{1}{4} (\lambda^2-1) \ .
\ee
It is generated by the modes
\cite{Feigin88,Pope:1989sr,Bordemann:1989zi, Bergshoeff:1989ns}
\be
V^s_n \ , \quad s\geq 2 \ , \quad |n|<s  \ ,
\ee
and contains an $\mathfrak{sl}(2)$ subalgebra generated by 
$V^2_{0,\pm 1}$ under which $V^s_n$ has spin $s-1$, 
\be\label{twoaction}
[V^2_m, V^s_n] = (-n + m(s-1)) V^s_{m+n} \  .
\ee
The bulk fields associated to $V^s_n$  then have spacetime spin $s$.  The full commutation relations are
of the form \cite{Pope:1989sr}
\be\label{vcom}
[V^s_m, V^t_n] = \sum_{ \stackrel{u=2}{\vspace{0.1cm} \mbox{\tiny even}}}^{s+t-1}  
g_u^{st}(m,n;\lambda) V_{m+n}^{s+t-u} \ ,
\ee
see for example \cite[Appendix A]{Gaberdiel:2011wb} for explicit formulae. The Lie algebra
\hs{\lambda} is a natural generalisation of $\mathfrak{sl}(N)$ to non-integer $N$
\cite{Feigin88,Fradkin:1990qk}. Indeed, for $\lambda=\pm N$, the modes $V^s_m$ with 
$s\geq N+1$ form an ideal $\chi_N$ in \hs{\lambda}, and 
\be\label{quotient}
\hs{\lambda=\pm N} / \chi_N \cong \mathfrak{sl}(N) \ .
\ee
This property played an important role in the arguments in Section~6 of \cite{GGHR}.
\smallskip

Given the structure of the commutation relations (\ref{vcom}) it is clear that the algebra
that is generated by the modes $V^s_m$ for which $s$ is even, form a subalgebra of \hs{\lambda}
\be
\hs{\lambda}^{(e)} \equiv \hbox{span} \{ V^s_m \in \hs{\lambda} \; : \; s \ \hbox{even} \} \ .
\ee
Note that $\hs{\lambda}^{(e)}$ contains in particular the $\mathfrak{sl}(2)$ algebra generated by 
$V^2_{0,\pm 1}$, and that the Chern-Simons theory based on it will lead to spin fields of all even 
spacetime spins.
This Chern-Simons theory is therefore a natural candidate for the higher spin theory that is dual
to the $\WD_N$ 't~Hooft limit. However, the situation is not as clear as for the case considered in
 \cite{Gaberdiel:2011wb,GGHR}. First, the analogue of  (\ref{quotient}) is now
\cite{Feigin88}
\be\label{quotient1}
\hs{\lambda=\pm N}^{(e)} / \chi_N = \left\{
\begin{array}{ll}
\mathfrak{sp}(N) \qquad & \hbox{if $N$ is even} \\
\mathfrak{so}(N) \qquad & \hbox{if $N$ is odd,}
\end{array}
\right.
\ee
and therefore does not lead to  $\mathfrak{so}$(even).\footnote{We thank Tom Hartman for very 
helpful comments on this point and the following discussion.} This does not necessarily disprove our 
suggestion since in relating bulk and boundary representations the Young tableaux have to be flipped,
see (\ref{3.34}),  and 
under this operation  $\mathfrak{sp}(-2N)$ is known to become $\mathfrak{so}(2N)$, see 
\cite{King,M,Cvitanovic:1982bq,MV}. 

Furthermore, we can think of the boundary algebra
as being the `unifying algebra' \cite{Blumenhagen:1994ik}  of the bulk description, since, at finite
$N$, the $\W_\infty$ algebra of the bulk Chern Simons theory should truncate to the relevant
boundary algebra. It follows from the analysis of  \cite[eq.\ (1.13) and (1.14)]{Blumenhagen:1994ik} 
that the  $\mathcal{WB}$-series has the ${\rm Orb}(\WD)$-series as unifying algebras, 
where 
`Orb' refers to the fact that the field of conformal dimension $h=N$ in $\WD_N$ 
has been removed --- this is probably the algebra that is relevant  for us since we do not consider any 
spinor representations and hence the field of conformal dimension $h=N$ is probably null.
This supports the idea that a $\mathcal{B}$-type higher spin algebra should underly the bulk description. 
However, we should mention that ${\rm Orb}(\WD_N)$ can {\em also} be the unifying
algebra of a $\mathcal{D}$-type higher spin algebra, see  \cite[eq.\ (1.8)]{Blumenhagen:1994ik},
and hence, again, the situation is not very clear cut.

The proposal also passes a simple consistency check: as in 
\cite{Gaberdiel:2011wb} the representations $(v;0)$ and $(0;v)$ of the $\WD_N$ 
't~Hooft limit correspond precisely to the irreducible $\mathfrak{sl}(2)$ representations
with highest weight $h=h_\pm$ that define irreducible representations of 
\hs{\lambda} because of  (\ref{Usl2}). It is easy to see from the explicit formula for the modes,
that these representations are also irreducible with respect to $\hs{\lambda}^{(e)}$, and
because of  (\ref{hvec}) the eigenvalues of $L_0$ agree, and it is implicit from 
(\ref{generalBra}) that the same is true for the characters. Finally, $\hs{\lambda}^{(e)}$ does 
not contain any generators of odd spin, and hence there are no separate conjugate representations. 
However, it is not clear how constraining these checks really are. 

In particular, there exists another natural possibility for the algebra of the higher spin Chern Simons
theory: \hs{\lambda} contains also the subalgebra
\be
\hso{\lambda} \equiv \hbox{span} \{ V^s_m \in \hs{\lambda} \; : \; s+m \ \hbox{even} \} \ ,
\ee
for which the analogue of (\ref{quotient1}) seems to be 
\be
\hso{\lambda=\pm N} / \chi_N = \mathfrak{so}(N)
\ee
for all $N$. It is therefore also possible that our $\WD_N$ 't~Hooft limit could be dual to the higher
spin theory based on \hso{\lambda}.  It would be very interesting to understand this issue in more detail,
for example, by repeating the analysis of \cite{Gaberdiel:2011wb} for 
$\hs{\lambda}^{(e)}$ and $\hso{\lambda}$. We should also get further insight into this problem
by studying the duality for the 't~Hooft limit of the $\W{\mathcal B}_N$ and $\W{\mathcal C}_N$ series 
in detail.  We hope to return to these questions elsewhere.

\section*{Acknowledgements}

This paper is based on the Master thesis of one of us (CV).
We thank Claude Eicher, Rajesh Gopakumar,  Misha Vasiliev, Roberto Volpato  and in particular
Tom Hartman for useful discussions and correspondences. The work of MRG is supported in part by 
the  Swiss National Science Foundation. 

\appendix

\section{Basics of $\mathfrak{so}(2N)$}\label{so}

In this appendix we review some basic properties of the simple Lie algebra $\mathfrak{so}(2N)$.  
The Cartan matrix $C$ of $\mathfrak{so}(2N)$  equals
\begin{equation}
C = \left( \begin{array}{cccccccc}
2 & -1 & 0 & 0 & \dots &  0 & 0 & 0\\
-1 & 2 & -1 & 0 & \dots &  0 & 0 & 0\\
0 & -1 & 2 & -1 & \dots &  0 & 0 & 0\\
0 & 0 & -1 & 2 & \dots &  0 & 0 & 0\\
. & . & . & . & \dots &  . & . & .\\
0 & 0 & 0 & 0 & \dots &  2 & - 1 & - 1\\
0 & 0 & 0 & 0 & \dots &  - 1 & 2 & 0 \\
0 & 0 & 0 & 0 & \dots &  - 1 & 0 & 2 \\
\end{array} \right) \ .
\end{equation}
For later use we also give the inverse Cartan matrix that defines the inner product
matrix of the fundamental weights
\begin{equation}\label{invC}
C^{-1} = 
\frac{1}{2} \left( \begin{array}{cccccccc}
2 & 2 & 2 & 2 & \dots &  2 & 1 & 1\\
2 & 4 & 4 & 4 & \dots &  4 & 2 & 2\\
2 & 4 & 6 & 6 & \dots &  6 & 3 & 3\\
2 & 4 & 6 & 8 & \dots &  8 & 4 & 4\\
. & . & . & . & \dots &  . & . & .\\
2 & 4 & 6 & 8 & \dots &  2 ( N - 2) & N - 2 & N -2\\
1 & 2 & 3 & 4 & \dots &  N - 2 & N / 2 & (N - 2) / 2 \\
1 & 2 & 3 & 4 & \dots &  N - 2 & (N - 2) / 2 & N / 2\\
\end{array} \right)\ .
\end{equation}
For many considerations it is convenient to work in an orthonormal basis 
\begin{equation}
\ve_1, \dots, \ve_N.
\end{equation}
With respect to this basis the simple roots of $\mathfrak{so}(2N)$ are given by
\begin{equation}
\begin{split}
\alpha_i &= \ve_i - \ve_{i+1}  \qquad (1 \leq i \leq N-1) \\
\alpha_N &= \ve_{N-1} + \ve_N\ .
\end{split}
\end{equation}
The fundamental weights are then
\begin{equation}
\begin{split}
\lambda_i &= \ve_1 + \ve_2 + \dots + \ve_{i}  \qquad  (1\leq i \leq N-2) \\
\lambda_{N-1} &= \tfrac{1}{2} \left( \ve_1 + \ve_2 + \dots + \ve_{N-1} - \ve_{N} \right) \\
\lambda_{N} &= \tfrac{1}{2} \left( \ve_1 + \ve_2 + \dots + \ve_{N-1} + \ve_{N} \right) \ ,
\end{split}
\end{equation}
and the Weyl vector equals
\begin{equation}\label{Weylv}
\rho = \sum_{i=1}^{N} \lambda_i =  \sum_{i = 1}^N (N -i) \ve_i.
\end{equation}
Its length square is therefore
\begin{equation}\label{Weylsq}
\rho^2 = \sum_{i = 1}^N (N -i)^2 = \frac{N (N-1) (2 N -1)}{6}\ .
\end{equation}

The highest weight $\la$ of a highest weight representation (hwr) can be expressed in terms
of the fundamental weights 
\begin{equation}
\la = \sum_{p = 1}^N \la_p \lambda_p\ , 
\end{equation}
where $\la_p\geq 0$ are the Dynkin labels of $\la$. (We shall usually write Dynkin labels
as $\Lambda=[\Lambda_1,\ldots,\Lambda_N]$.) Sometimes it is also convenient to 
expand $\la$ with respect to the above orthonormal basis as
\begin{equation}\label{LamONB}
\la = \sum_{p = 1}^N l_i \ve_i\ .
\end{equation}
These expansion coefficients are related to the Dynkin labels via
\begin{equation}\label{lidef}
\begin{split}
l_i &= \sum_{p = i}^{N-2} \la_p + \tfrac{1}{2} \left( \la_{N-1} + \la_{N} \right) \qquad (1\leq i \leq N-2) \\
l_{N-1} &= \tfrac{1}{2} \left( \la_{N-1} + \la_{N} \right) \ , \qquad l_{N} = \tfrac{1}{2} \left( \la_{N} - \la_{N-1} \right)\ .
\end{split}
\end{equation}
In the hwr $\la$, the quadratic Casimir takes the value
\begin{equation}\label{quadC}
C_N(\la) = \frac{1}{2} \left\langle \la, \la + 2 \rho \right\rangle 
= \frac{1}{2} \sum_{i = 1}^N l_i^2 + \sum_{i = 1}^N l_i \, (N-i) \ .
\end{equation}
For example, for the vector representation whose Dynkin labels are 
$v=[1,0^{N-1}]$, the value of the quadratic Casimir $C_N$ equals
\begin{equation}\label{A.12}
C_N (v) = \frac{1}{2} \left \langle \Lambda_v, \Lambda_v + 2 \rho \right \rangle = 
\frac{1}{2} + (N-1) = \frac{2 N - 1}{2}\ ,
\end{equation}
while for the two spinor representations $s=[0^{N-2},1,0]$ and $c=[0^{N-1},1]$ we have
\begin{equation}\label{A.13}
C_N(s) = C_N(c) = \frac{N}{8} + \frac{N (N-1)}{4} =  \frac{N(2N-1)}{8} \ .
\end{equation}
Note that the quadratic Casimir of the vector representation $v$ is proportional to $N$, while that of the
two spinor representations $s$ and $c$ is proportional to $N^2$. This last property is true for
any representation in either of the two spinor conjugacy classes. To see this, 
suppose that a hwr $\la = [\la_1, \dots, \la_N]$ has the property that $a\equiv \max(\la_{N-1},\la_N)>0$. 
Then  
\be
\langle \la,\la\rangle \geq  \frac{a^2}{4}\, N \ , \qquad
\langle \la, \rho \rangle \geq \frac{a}{4} N (N-1) \ .
\ee
Both of these claims follow directly from the form of the inverse Cartan matrix $C^{-1}$, see (\ref{invC}):
since all its entries are positive, we have
\begin{eqnarray}\label{A.15}
\langle \la,\la\rangle & \geq & \langle [0,0,\ldots, 0 ,a], [0,0,\ldots,0,a]\rangle = \frac{a^2}{4}\, N \\
\langle \la, \rho \rangle & \geq & \langle [0,0,\ldots,0,a],[1,1,\ldots,1]\rangle =
\frac{a}{2} \left(\sum_{i=1}^{N-2} i  + \tfrac{N-2}{2} + \tfrac{N}{2} \right) 
= \frac{a}{4} N (N-1) \ . \nonumber 
\end{eqnarray}
Note that (\ref{A.15}) implies immediately that the quadratic Casimir of any representation with 
$\max(\la_{N-1},\la_N)>0$ is proportional to $N^2$. Actually, the same will be true for any
representation for which $\Lambda_{N-j}\neq 0$ for any fixed $j$. 

For most of the paper we will only be interested in representations for which the quadratic 
Casmir grows linearly with $N$. These representations, in particular, satisfy
$\la_{N-1}=\la_N=0$, and it then follows from (\ref{lidef}) that they 
are labelled by $l_i\in \NN$, with $l_i\geq l_{i+1}$ for $i=1,\ldots, N-2$, 
and $l_{N-1}=l_{N}= 0$, see (\ref{lidef}). We can thus think of $l_i$ as the number of boxes in the
$i^{\rm th}$ row, and hence label these representations by Young tableaux. In terms of these
Young tableaux, let us also denote the number of boxes in the $j^{\rm th}$ column by $c_j$.
Then we have the identity
\begin{eqnarray}
\tfrac{1}{2} \sum_j c_j^2 & = &  \tfrac{1}{2} \sum_{j}^{N} (l_j - l_{j+1}) j^2 =
\tfrac{1}{2 } \sum_{j = 1}^N l_j \, j^2 - \tfrac{1}{2} \sum_{j = 1}^{N-1} l_{j+1} \big( (j+1)^2 - 2j - 1 \big) 
\nonumber \\
& = & \tfrac{1}{2} l_1 + \sum_{j = 1}^{N-1} l_{j+1} \, j  + \tfrac{1}{2} \sum_{j = 1}^{N-1} l_{j+1}  
= \sum_{j = 1}^{N} l_j \, j - \tfrac{1}{2} \sum_{j = 1}^{N} l_j \ ,
\end{eqnarray}
from which we deduce that the quadratic Casimir of a hwr $\la$ with $\la_{N-1}=\la_N=0$ equals
\begin{equation}\label{A.18}
C_N(\la) = \tfrac{1}{2} \left\langle \la, \la + 2 \rho \right\rangle 
= \tfrac{1}{2} \sum_{i = 1}^N l_i^2 + \sum_{i = 1}^N l_i \, (N-i)
= B (\la) \left(N - \tfrac{1}{2} \right) + \tfrac{1}{2} D(\la),
\end{equation}
where
\begin{equation}\label{A.17}
B(\la) = \sum_{i = 1}^{N} l_i \ , \qquad D(\la) = \sum_{i=1}^N l_i^2 - \sum_{i = 1}^N c_i^2 \ . 
\end{equation}

\subsection{The Weyl denominator formula}

The Weyl denominator formula states (see {\it e.g.} \cite[eq.\ (1.7.39)]{Fuchs}) 
\begin{equation}\label{Weyl}
\sum_{w \in W} \epsilon(w) e^{\langle w(\rho),h \rangle} 
= e^{\langle \rho, h \rangle} \prod_{\alpha > 0} (1 - e^{- \langle \alpha, h \rangle} ) \ ,
\end{equation}
where the sum is over the Weyl group $W$ of the finite dimensional Lie algebra $\mathfrak{g}$, while
the product is over all positive roots of $\mathfrak{g}$. If we set $h = - \xi \rho$, where $\xi \in \mathbb{C}$
and $\rho$ is the Weyl vector, and define $q \equiv e^{\xi}$, then (\ref{Weyl}) becomes
\begin{equation}
\sum_{w \in W} \epsilon(w) q^{ - \langle w(\rho), \rho \rangle} 
= q^{- \rho^2} \prod_{\alpha > 0} (1 - q^{\langle \alpha, \rho  \rangle})\ .
\end{equation}
For the case of $\mathfrak{g} = \mathfrak{so}(2N)$, the positive roots are of the form 
\begin{equation}\label{positive}
\varepsilon_i - \varepsilon_{i+\ell} \quad \mathrm{and} \quad \varepsilon_i + \varepsilon_{i+\ell} 
\qquad (\ell >0) \ .
\end{equation}
For the former, the inner product with the Weyl vector equals 
\begin{equation}
\langle \varepsilon_i - \varepsilon_{i+\ell} , \rho \rangle = \ell \ ,
\end{equation}
and for fixed $\ell$ with $1\leq \ell \leq N-1$, there are $(N-\ell)$ positive roots whose inner product
equals $\ell$. For the other class of positive roots we have instead
\begin{equation}\label{2in}
\langle \varepsilon_i + \varepsilon_{i+\ell}, \rho \rangle = 2 (N - i) - \ell \equiv m \ .
\end{equation}
For $1 \leq m \leq N-1$ there are exactly $\left \lfloor \frac{m + 1}{2} \right \rfloor$ roots
of the form $\varepsilon_i + \varepsilon_{i+\ell}$ for which (\ref{2in}) equals $m$, while for 
$N\leq m \leq 2N-3$ their number equals $N - \left \lceil \frac{m + 1}{2} \right \rceil$. (Here
$\left \lfloor x \right \rfloor$ is the largest integer less or equal than $x$, while 
$\left \lceil x \right \rceil$ is the smallest integer bigger or equal than $x$.) Combining these
results we thus conclude that the Weyl denominator formula for $\mathfrak{so}(2N)$ is of the
form 
\begin{equation}\label{Weylso}
\sum_{w \in W} \epsilon(w) q^{ - \langle w(\rho), \rho \rangle} = 
q^{- \rho^2} \prod_{n=1}^{2N-3} (1 - q^n)^{N- \left \lceil \frac{n - 1}{2} \right \rceil} 
\prod_{n=N}^{2N-3} (1 - q^n)^{-1} \ . 
\end{equation}
\medskip

\noindent We can similarly use (\ref{Weyl}) to rewrite
\begin{equation} \label{A.24}
\sum_{w \in W} \epsilon(w) q^{ - \langle w(\Lambda + \rho), \rho \rangle} 
= \sum_{w \in W} \epsilon(w) q^{ - \langle w(\rho), \Lambda + \rho \rangle}
= q^{-{\rho}^2 - \langle \Lambda, \rho \rangle} \prod_{\alpha > 0} 
\left( 1- q^{\langle \alpha, \Lambda + \rho \rangle} \right)\ .
\end{equation}
Working in the orthonormal basis, {\it i.e.}\ using (\ref{LamONB}) and (\ref{Weylv}),
we obtain for the two different classes of positive roots (\ref{positive})
\begin{eqnarray}
\langle \varepsilon_j - \varepsilon_i, \Lambda + \rho \rangle & = & 
\bigl\langle \varepsilon_j - \varepsilon_i, 
\sum_{m = 1}^{N} l_m\, \varepsilon_m + \sum_{m = 1}^N (N -m) \ve_m  \bigr \rangle = l_j - l_i + i - j \nonumber \\
\langle \varepsilon_j + \varepsilon_i, \Lambda + \rho \rangle & = & 
\bigl\langle \varepsilon_j + \varepsilon_i, 
\sum_{m = 1}^{N} l_m\, \varepsilon_m + \sum_{m = 1}^N (N -m) \ve_m  \bigr \rangle 
= l_j + l_i + 2N - i - j \ , \nonumber
\end{eqnarray}
where $i>j$.  Hence (\ref{A.24}) becomes
\begin{equation} \label{nom}
\sum_{w \in W} \epsilon(w) q^{ - \langle w(\Lambda + \rho), \rho \rangle}
= q^{-{\rho}^2 - \langle \Lambda, \rho \rangle} 
\prod_{i = 2}^N \prod_{j = 1}^{i - 1} \left( 1 - q^{l_j - l_i + i - j} \right)
\left( 1 - q^{l_j + l_i + 2N - i - j} \right) \ ,
\end{equation}
which for $\la = 0$ reduces to 
\begin{equation} \label{denom}
\sum_{w \in W} \epsilon(w) 
q^{ - \langle w(\rho), \rho \rangle} = 
q^{-{\rho}^2} \prod_{i = 2}^N \prod_{j = 1}^{i - 1} \left( 1 - q^{i - j} \right)
\left( 1 - q^{2N - i - j} \right)\ .
\end{equation}

\section{The fusion calculation}\label{fusion}

In this appendix we give some details of the fusion calculation $(v;v)\equiv (v;0)\otimes (0;v)$. Since
most of the steps are the same as in the calculation of \cite{GGHR}, we shall be brief. 

\noindent The two representations $\phi_1\equiv (v;0)$ and $\phi_2\equiv (0;v)$ have the 
eigenvalues
\begin{eqnarray}
\phi_1\equiv (v;0) : & \qquad  & 
h_1 = \tfrac{1}{2} (1+\lambda) \ , \qquad 
u_1 = (1+\lambda) (2+\lambda) (3+\lambda)  \nonumber \\
\phi_2\equiv (0;v) : & \qquad  & 
h_2 = \tfrac{1}{2} (1-\lambda) \ , \qquad 
u_2 = (1-\lambda) (2-\lambda) (3-\lambda)  \ ,
\end{eqnarray}
and we have the null vectors
\be
\begin{array}{ll}
U_{-1} \phi_1 = 4 (2+\lambda) (3+\lambda) L_{-1} \phi_1 \qquad 
&U_{-1} \phi_2 = 4 (2-\lambda) (3-\lambda) L_{-1} \phi_2  \\
U_{-2} \phi_1 = 10 (3+\lambda) L_{-1}^2 \phi_1 \qquad 
&U_{-2} \phi_2 = 10 (3-\lambda) L_{-1}^2 \phi_2 \\
U_{-3} \phi_1 = 20 L_{-1}^3 \phi_1 \qquad 
&U_{-3} \phi_2 = 20 L_{-1}^3 \phi_2 \ .
\end{array}
\ee
For the calculation of the highest weight space we can use the relations
\be\label{eq1}
\begin{array}{ll}
L_{-1}: \qquad & (L_{-1}\otimes {\bf 1}) + ({\bf 1} \otimes L_{-1}) \cong 0  \\
U_{-1}: \qquad & (U_{-3}\otimes {\bf 1}) + 2 (U_{-2}\otimes {\bf 1}) + (U_{-1}\otimes {\bf 1}) 
+ ({\bf 1} \otimes U_{-1}) \cong 0  \\
\qquad \quad\Rightarrow
& 20 (L_{-1}^3\otimes {\bf 1}) + 20 (3+\lambda) (L_{-1}^2\otimes {\bf 1}) 
+ 40 \lambda (L_{-1}\otimes {\bf 1}) \cong 0 \ ,
\end{array}
\ee
where $(S_1\otimes S_2)\equiv (S_1\phi_1 \otimes S_2\phi_2)$, and we have used the null relations,
as well as  the $L_{-1}$ relation in the last line.  Furthermore, we get from $U_{-2}$
\be\label{eq2}
20 (L_{-1}^3 \otimes {\bf 1}) + 60 (L_{-1}^2 \otimes {\bf 1}) \cong 0 \ ,
\ee
where we have used that 
\be
(L_{-1}^2\otimes {\bf 1}) \cong - (L_{-1} \otimes L_{-1}) \cong ({\bf 1} \otimes L_{-1}^2) \ ,
\ee
which follows from the $L_{-1}$ relation.  Combining (\ref{eq1}) and (\ref{eq2}) we conclude that
\be
20 \lambda \Bigl[ (L_{-1}^2\otimes {\bf 1}) + 2  (L_{-1}\otimes {\bf 1}) \Bigr] \cong 0 \ .
\ee
Provided that $\lambda\neq 0$, we thus have 
$(L_{-1}^2\otimes {\bf 1}) + 2  (L_{-1}\otimes {\bf 1}) \cong 0$, which reproduces precisely
eq.\ (5.17) of \cite{GGHR} (which was there derived using the $W$-modes which we now
do not have at our disposal). The rest of the analysis then proceeds exactly as in \cite{GGHR}.

\subsection{Going up to level one}

At level one we still have all but the first relation from \cite[eq.\ (B.1)]{GGHR} 
\be\label{eq3}
\begin{array}{lrl}
U_{-3}: \quad & 
(L_{-1}^3 \otimes {\bf 1}) & \cong - ({\bf 1} \otimes  L_{-1}^3) \\
U_{-2}: \quad & 
(L_{-1}^3 \otimes {\bf 1}) & \cong - \tfrac{1}{2} (3+\lambda) (L_{-1}^2 \otimes {\bf 1})
- \tfrac{1}{2}  (3-\lambda)  ({\bf 1} \otimes  L_{-1}^2) \\
L_{-1} L_{-1}: \quad 
& (L_{-1}^2 \otimes {\bf 1}) & \cong  - 2 (L_{-1} \otimes L_{-1}) - 
({\bf 1} \otimes L_{-1}^2) \ .
\end{array}
\ee
In order to obtain the missing first relation
we consider now the identity coming from $L_{-1} U_{-1}$, which leads to 
\begin{eqnarray}
0 & \cong & 20 \bigl[ (L_{1}^4\otimes {\bf 1}) + (L_{-1}^3 \otimes L_{-1}) \bigr] 
+ 20 \, (3+\lambda) \bigl[ (L_{-1}^3 \otimes {\bf 1}) +  (L_{-1}^2 \otimes L_{-1})\bigr] \nonumber \\
& & +4(2+\lambda) (3+\lambda) \bigl[ (L_{-1}^2 \otimes {\bf 1}) + (L_{-1} \otimes L_{-1}) \bigr] \nonumber \\
& & + 4(2-\lambda) (3-\lambda) \bigl[  (L_{-1} \otimes L_{-1})  + ({\bf 1} \otimes L_{-1}^2 ) \bigr] \ ,
 \end{eqnarray}
while $L_{-1}U_{-2}$ gives
\be
(L_{1}^4\otimes {\bf 1}) + (L_{-1}^3 \otimes L_{-1})  \cong 
- \lambda (L_{-1}^3\otimes {\bf 1}) - \tfrac{1}{2} (3+\lambda)  (L_{-1}^2 \otimes L_{-1}) 
- \tfrac{1}{2} (3-\lambda)  (L_{-1} \otimes L_{-1}^2)  \ .
\ee
Combining these equations we  deduce 
\be\label{eq4}
0 \cong 6 (L_{-1}^3 \otimes {\bf 1}) + (3+\lambda) (L_{-1}^2 \otimes L_{-1})
- (3-\lambda) (L_{-1} \otimes L_{-1}^2) 
+ 2 \lambda  \bigl[ (L_{-1}^2 \otimes {\bf 1}) - ({\bf 1}\otimes L_{-1}^2) \bigr] \ .
\ee
Next we use the relation coming from $L_{-1}^3$, together with the first relation from
(\ref{eq3}), to get 
\be\label{eq5}
0 \cong 
(L_{-1}^2 \otimes L_{-1}) + (L_{-1} \otimes L_{-1}^2) \ .
\ee
Combining this with (\ref{eq4}) this leads to 
\be\label{eq6}
0 \cong 6  (L_{-1}^3 \otimes {\bf 1}) + 6  (L_{-1}^2 \otimes L_{-1})  
+ 2 \lambda  \bigl[ (L_{-1}^2 \otimes {\bf 1}) - ({\bf 1}\otimes L_{-1}^2) \bigr] \ .
\ee
From $L_{-1}^2$ applied to $(L_{-1}\otimes {\bf 1})$ we obtain
\be
0 \cong (L_{-1}^3\otimes {\bf 1}) + 2 (L_{-1}^2 \otimes L_{-1}) + (L_{-1} \otimes L_{-1}^2) 
\ee
which, together with (\ref{eq5}), leads to 
\be
0 \cong (L_{-1}^3\otimes {\bf 1}) + (L_{-1}^2 \otimes L_{-1}) \ .
\ee
Together with (\ref{eq6}) this then implies
\be
0 \cong 2 \lambda  \bigl[ (L_{-1}^2 \otimes {\bf 1}) - ({\bf 1}\otimes L_{-1}^2) \bigr] \ ,
\ee
from which we conclude, for $\lambda\neq 0$, the missing first relation of  \cite[eq.\ (B.1)]{GGHR} 
\be
(L_{-1}^2 \otimes {\bf 1}) \cong ({\bf 1}\otimes L_{-1}^2) \ .
\ee
The rest of the analysis is identical to \cite[Appendix B.1]{GGHR}, and thus the structure of the
resulting representation is as described there.


\begin{thebibliography}{99}

\bibitem{Prokushkin:1998bq}
S.F.~Prokushkin and M.A.~Vasiliev,
``Higher spin gauge interactions for massive matter fields in 3d AdS space-time,"
Nucl.\ Phys.\  B {\bf 545} (1999) 385
{\tt [arXiv:hep-th/9806236]}.

\bibitem{Prokushkin:1998vn}
S.~Prokushkin and M.A.~Vasiliev,
``3-d higher spin gauge theories with matter,''
{\tt arXiv:hep-th/9812242}.

\bibitem{Gaberdiel:2010pz}
M.R.~Gaberdiel and R.~Gopakumar,
``An AdS$_3$ dual for minimal model CFTs,''
Phys.\ Rev.\  D {\bf 83} (2011) 066007 
{\tt  [arXiv:1011.2986 [hep-th]]}.

 \bibitem{Klebanov:2002ja}
 I.R.~Klebanov and A.M.~Polyakov,
``AdS dual of the critical O(N) vector model,"
Phys.\ Lett.\  B {\bf 550} (2002) 213
{\tt [arXiv:hep-th/0210114]}.

\bibitem{Witten}
E.~Witten, talk at the John Schwarz 60-th birthday symposium, \newline
{\tt http://theory.caltech.edu/jhs60/witten/1.html}

\bibitem{Mikhailov:2002bp}
A.~Mikhailov,
``Notes on higher spin symmetries,''
{\tt arXiv:hep-th/0201019}.

\bibitem{Sezgin:2002rt}
E.~Sezgin and P.~Sundell,
``Massless higher spins and holography,"
Nucl.\ Phys.\  B {\bf 644} (2002) 303
[Erratum-ibid.\  B {\bf 660} (2003) 403]
{\tt [arXiv:hep-th/0205131]}.

\bibitem{Vasiliev:2003ev}
M.A.~Vasiliev,
``Nonlinear equations for symmetric massless higher spin fields in (A)dS(d),"
Phys.\ Lett.\  B {\bf 567} (2003)  139
{\tt [arXiv:hep-th/0304049]}.

\bibitem{Vasiliev:1999ba}
M.A.~Vasiliev,
``Higher spin gauge theories: Star product and AdS space,"
in M.A.~Shifman (ed.), `The many faces of the superworld', p.\ 533
{\tt [arXiv:hep-th/9910096]}.

\bibitem{Bekaert:2005vh}
X.~Bekaert, S.~Cnockaert, C.~Iazeolla and M.A.~Vasiliev,
``Nonlinear higher spin theories in various dimensions,"
{\tt arXiv:hep-th/0503128}.

\bibitem{Iazeolla:2008bp}
C.~Iazeolla,
``On the algebraic structure of higher-spin field equations and new exact solutions,''
{\tt arXiv:0807.0406 [hep-th]}.

\bibitem{Campoleoni:2009je}
A.~Campoleoni,
``Metric-like Lagrangian formulations for higher-spin fields of mixed symmetry,"
Riv.\ Nuovo Cim.\  {\bf 033} (2010) 123
{\tt [arXiv:0910.3155 [hep-th]]}.

\bibitem{Giombi:2009wh}
S.~Giombi and X.~Yin,
``Higher spin gauge theory and holography: the three-point functions,"
JHEP {\bf  1009} (2010) 115 
{\tt [arXiv:0912.3462 [hep-th]]}.
 
\bibitem{Giombi:2010vg}
S.~Giombi and X.~Yin,
``Higher spins in AdS and twistorial holography," 
JHEP {\bf 1104} (2011) 086
{\tt [arXiv:1004.3736 [hep-th]]}.

\bibitem{Koch:2010cy}
R.d.M.~Koch, A.~Jevicki, K.~Jin and J.P.~Rodrigues,
``AdS$_4$/CFT$_3$ construction from collective fields,"
Phys.\ Rev.\ D {\bf 83} (2011) 025006
{\tt [arXiv:1008.0633 [hep-th]]}.
 
 \bibitem{Giombi:2011ya}
S.~Giombi and X.~Yin,
``On higher spin gauge theory and the critical O(N) model,''
{\tt arXiv:1105.4011 [hep-th]}.

\bibitem{Vasiliev:1992ix}
M.A.~Vasiliev,
``Equations of motion for d = 3 massless fields interacting through 
Chern-Simons higher spin gauge fields,''
Mod.\ Phys.\ Lett.\  A {\bf 7} (1992) 3689.

\bibitem{Bouwknegt:1992wg}
P.~Bouwknegt and K.~Schoutens,
``W symmetry in conformal field theory,"
Phys.\ Rept.\  {\bf 223} (1993) 183
{\tt [arXiv:hep-th/9210010]}.
 
\bibitem{GGHR}
M.R.~Gaberdiel, R.~Gopakumar, T.~Hartman and S.~Raju,
``Partition functions of holographic minimal models,"
{\tt arXiv:1106.1897 [hep-th]}.

\bibitem{Gaberdiel:2011wb}
M.R.~Gaberdiel and T.~Hartman,
``Symmetries of holographic minimal models,''
JHEP {\bf 1105} (2011) 031 
{\tt [arXiv:1101.2910 [hep-th]]}.  

\bibitem{Henneaux:2010xg}
M.~Henneaux and S.-J.~Rey,
``Nonlinear $W_{\infty}$ as asymptotic symmetry of three-dimensional higher spin 
anti-de Sitter gravity,''
JHEP {\bf 1012} (2010) 007 \newline
{\tt [arXiv:1008.4579 [hep-th]]}.

\bibitem{Campoleoni:2010zq}
 A.~Campoleoni, S.~Fredenhagen, S.~Pfenninger and S.~Theisen,
``Asymptotic symmetries of three-dimensional gravity coupled to higher-spin fields,''
JHEP {\bf 1011} (2010) 007
{\tt  [arXiv:1008.4744 [hep-th]]}.

\bibitem{Gaberdiel:2010ar}
M.R.~Gaberdiel, R.~Gopakumar and A.~Saha,
``Quantum $W$-symmetry in AdS$_3$,''
JHEP {\bf 1102} (2011) 004
{\tt  [arXiv:1009.6087 [hep-th]]}.

\bibitem{David:2009xg}
J.R.~David, M.R.~Gaberdiel and R.~Gopakumar,
``The heat kernel on AdS$_3$ and its applications,"
JHEP {\bf 1004} (2010) 125 
{\tt [arXiv:0911.5085 [hep-th]]}.
 
\bibitem{Giombi:2008vd}
S.~Giombi, A.~Maloney and X.~Yin,
``One-loop partition functions of 3D gravity,''
JHEP {\bf 0808} (2008) 007 
{\tt [arXiv:0804.1773 [hep-th]]}.

\bibitem{Fuchs}
J.~Fuchs, 
``Affine Lie algebras and quantum groups," Cambridge University Press
(1992). 

\bibitem{MacDonald}
I.G.~MacDonald, 
``Symmetric functions and Hall polynomials," Oxford University Press (1979). 

\bibitem{Feigin88} 
B.L.~Feigin, 
``Lie algebras gl$(\lambda)$ and cohomologies of  Lie algebras of 
differential operators," 
Russian Mathematical Surveys {\bf 43} no.\ 2 (1988) 169.

\bibitem{Pope:1989sr}
C.N.~Pope, L.J.~Romans and X.~Shen,
``W($\infty$) and the Racah-Wigner algebra,"
Nucl.\ Phys.\  B {\bf 339} (1990) 191.

\bibitem{Bordemann:1989zi}
M.~Bordemann, J.~Hoppe and P.~Schaller,
``Infinite dimensional matrix algebras,"
Phys.\ Lett.\  B {\bf 232}  (1989) 199.

\bibitem{Bergshoeff:1989ns}
E.~Bergshoeff, M.P.~Blencowe and K.S.~Stelle,
``Area preserving diffeomorphisms and higher spin algebra,''
Commun.\ Math.\ Phys.\  {\bf 128}  (1990) 213.

\bibitem{Fradkin:1990qk}
E.S.~Fradkin and V.Y.~Linetsky,
``Supersymmetric Racah basis, family of infinite dimensional superalgebras, 
SU($\infty$ + 1$|\infty$) and related 2-D models,''
Mod.\ Phys.\ Lett.\  A {\bf 6} (1991) 617.

\bibitem{King}
R.C.~King,
``The dimensions of irreducible tensor representations of the orthogonal and symplectic groups,"
Can.\ J.\  Math.\ {\bf 33} (1972) 176.

\bibitem{M}
R.L.~Mkrtchyan,
``The equivalence of Sp(2N) and SO(-2N) gauge theories,"
Phys.\ Lett.\ B {\bf 105} (1981) 174.

\bibitem{Cvitanovic:1982bq}
P.~Cvitanovic and A.~D.~Kennedy,
``Spinors in negative dimensions,"
Phys.\ Scripta {\bf 26} (1982) 5.

\bibitem{MV}
R.L.~Mkrtchyan and  A.P.~Veselov,
``On duality and negative dimensions in the theory of Lie groups and symmetric spaces,"
{\tt arXive:1011.0151 [math.RT]}.

\bibitem{Blumenhagen:1994ik}
R.~Blumenhagen, W.~Eholzer, A.~Honecker, K.~Hornfeck and R.~Hubel,
``Unifying W algebras,''
Phys.\ Lett.\  B {\bf 332} (1994) 51
{\tt [arXiv:hep-th/9404113]}.

\bibitem{Ahn:2011pv}
C.~Ahn,
``The large N 't Hooft limit of coset minimal models,''
{\tt arXiv:1106.0351 [hep-th]}.
 
 
 
\end{thebibliography}
\end{document}